\documentclass[preprint,12pt]{elsarticle}
\usepackage{txfonts}
\usepackage{graphicx}% Include figure files
\usepackage{dcolumn}% Align table columns on decimal point
\usepackage{amssymb}
\usepackage{color}

\newcommand{\beq}{\begin{equation}}
\newcommand{\eeq}{\end{equation}}
\newcommand{\beqa}{\begin{eqnarray}}
\newcommand{\eeqa}{\end{eqnarray}}

\journal
{Physics Letters B}

\begin{document}

\begin{frontmatter}
\title{PAMELA/Fermi-LAT electron cosmic ray spectrum at $\sim$100 GeV:
implication for dark matter annihilation signal in accordance with the
130 GeV $\gamma$-ray line}

\author{Lei Feng$^{a}$, Qiang Yuan$^{b,a}$, Xiang Li$^{a}$, and Yi-Zhong Fan$^{a}$\footnote{The corresponding author (email: yzfan@pmo.ac.cn)}}

\address[a] {Key Laboratory of Dark Matter and Space Astronomy, Purple Mountain Observatory, Chinese Academy of Sciences, Nanjing
210008, China}
\address[b]{Key Laboratory of Particle Astrophysics, Institute of High Energy
Physics, Chinese Academy of Science, Beijing 100049, China
\\}
\begin{abstract}
Recently, a tentative 130 GeV $\gamma$-ray line signal was identified
by quite a few groups. If correct it would constitute a ``smoking gun''
for dark matter annihilations. Interestingly, the spectra of the
cosmic ray electrons detected by PAMELA and Fermi-LAT both show tiny
wiggle-like structure at $\sim 100$ GeV, which might indicate a weak
signal of the annihilation of $\sim 130$ GeV dark matter particles into
electrons/positrons with a velocity-weighted cross section
$\langle\sigma v\rangle_{\rm \chi\chi\rightarrow e^{+}e^{-}}
\sim 4\times10^{-26}~{\rm cm^{3}~s^{-1}}$. The prospect of identifying
such a potential weak dark-matter-annihilation electron and/or positron
component by AMS-02, a mission to measure the high energy cosmic ray
spectra with unprecedented accuracy, is investigated.
\end{abstract}
\end{frontmatter}

\section{Introduction}
Gamma-ray line is generally thought to be a smoking gun observation of
dark matter (DM). Recently, Bringmann {\it et al.} \cite{bringman} and
Weniger \cite{weniger} reported that there might be hint of a
monochromatic $\gamma$-ray line with energy $\sim130$ GeV in the
data recorded by Fermi Large Area Telescope (Fermi-LAT) \cite{fermi}.
This $\gamma$-ray line could be explained by $\sim 130$ GeV DM
particle annihilation, with the velocity-weighted cross section
$\langle\sigma v\rangle_{\rm \chi\chi\rightarrow \gamma\gamma}
\sim 10^{-27} {\rm cm^3~s^{-1}}$. This phenomenon was confirmed by
a series of independent analyses \cite{tempel,boyarsky,sumeng}.
It was argued that such a line-like structure might originate from
astrophysical emission related with the Fermi bubbles \cite{profumo}
but the morphology analysis indicated that it is not the case \cite{tempel}.
Based on the identified spectral and spatial variations of rich
structures of the diffuse $\gamma$-ray emission in the inner Galaxy,
Boyarsky {\it et. al.} argued against the DM origin of these
structures \cite{boyarsky}. However, the DM origin of the
$\gamma$-ray line emission has been strengthened by Su \& Finkbeiner
\cite{sumeng} and by Yang et al. \cite{Yang2012}. The independent
analyses to search for $\gamma$-ray
lines in the Milky Way halo by Fermi-LAT collaboration
\cite{fermiline} and in dwarf galaxies by \cite{alex} found no
significant signal, but the constraints are not tight enough to
exclude such a $\gamma$-ray line signal. It was also proposed that
such a line-like signal could be tested with high energy resolution
detectors in the near future \cite{test}. Several models had been
proposed to explain this tentative line structure \cite{model1}.

Several years ago ATIC experiment discovered significant excess in
the $e^+ + e^-$ energy spectrum between $300-800$ GeV,
moreover the $e^+ + e^-$ energy spectrum also showed possible wiggle-like
structure at $\sim 100$ GeV \cite{atic}, which has been studied
by a few research groups \cite{twocomponent}.
The $e^-$ spectra measured by PAMELA and Fermi-LAT both revealed
tentative fine structure above $\sim100$ GeV \cite{pamela,Fermi2012}.
Therefore a natural question one would ask is whether there is any
connection between the $130$ GeV line-like structure of $\gamma$-rays
and the wiggle structure of electrons.

In this Letter, we focus on the PAMELA/Fermi-LAT electron data and show
that the DM scenario with mass $\sim 130$ GeV corresponding to the
possible $\gamma$-ray line, might contribute to the tentative fine
structure of the $e^-$ spectra around $100$ GeV.
Together with the PAMELA/Fermi-LAT positron fraction
\cite{Fermi2012,pamela-positron} data, we set a constraint on the
velocity-averaged cross section $\langle\sigma v\rangle_{\rm
\chi\chi\rightarrow e^{+}e^{-}} < 10^{-25}$ cm$^3$ s$^{-1}$, consistent
with all of the current bounds of the indirect detection measurements.
Considering the large uncertainties of the current data, more advanced
and dedicated experimental observations, in particular by
AMS-02\footnote{http://www.ams02.org/}, are highly necessary to pin
down the shape of the electron and positron spectrum, and then confirm
or rule out a DM component in accordance with the $\sim 130$ GeV
$\gamma-$ray line.

\section{Cosmic ray propagation}

The cosmic ray (CR) propagation equation is written as follows
\cite{galprop}:
\begin{eqnarray}
\frac{\partial\psi}{\partial
t}&=&q({\bf r},p)+\nabla\cdot\left(D_{xx}\nabla\psi-{\bf V}\psi\right)+
\frac{\partial}{\partial p}p^{2}D_{pp}\frac{\partial}{\partial p}
\frac{\psi}{p^{2}} \nonumber \\
 &-& \frac{\partial}{\partial p}\left[\dot{p}\psi-\frac{p}{3}(\nabla\cdot
{\bf V})\psi\right]-\frac{\psi}{\tau_{f}}-\frac{\psi}{\tau_{r}},
\end{eqnarray}
where $\psi=\psi({\bf r},p,t)$ is the density per unit of total particle
momentum, $q({\bf r},p)$ is the source distribution function, $D_{xx}$
is the spatial diffusion coefficient, ${\bf V}=dV/dz\times z$ is the
convection velocity, $D_{pp}$ is the diffusion coefficient in momentum
space, $\dot{p}=dp/dt$ is the momentum loss rate, $\tau_f$ and $\tau_r$
are the time scales of fragmentation and radioactive decay.

In general it is difficult to solve the propagation equation with
analytical method, given the complicated distributions of the
source, interstellar matter, radiation field and magnetic field.
Numerical methods are developed to solve the propagation equations,
such as GALPROP \cite{galprop} and DRAGON \cite{dragon}. In this
work we adopt the GALPROP package to calculate the propagation of
the CR particles, including the contribution from DM annihilation.
The diffusion-reacceleration (DR) and diffusion-convection (DC) models
of CR propagation are adopted as illustration. The main parameters of
propagation and source injection are compiled in Table. \ref{table1}.
These set of propagation parameters can fit the observational B/C,
$^{10}$Be/$^9$Be and proton data \cite{ypf}.

\begin{table}[!htb]
\begin{small}
\caption {The propagation parameters.}
\begin{tabular}{cccccccc}
\hline \hline
& $z_{h}$ & $D_{0}$ & diffusion index\footnotemark[1] & $v_A$ &$dV_{c}/dz$ & $e^-$ injection\footnotemark[2] & $E_{\rm br}$\\
& (kpc) & ($10^{28}$ cm$^2$ s$^{-1}$) & $\delta_1/\delta_2$  & (km s$^{-1}$)  & (km s$^{-1}$ kpc$^{-1}$)& $\gamma_1/\gamma_2$ & (GeV)\\
\hline
DR & $3.9$ & $6.6$ & $0.30/0.30$ & $39.2$  & 0 & $1.61/2.70$ & $4.3$ \\
DC & $3.9$ & $2.5$ & $0/0.55$ & 0 &$6$ & $1.63/2.74$ & $4.0$ \\
\hline
\hline
\end{tabular}
\label{table1}
\footnotemark[1]{Below/above rigidity $\rho_0=4$ GV.} \\
\footnotemark[2]{Below/above $E_{\rm br}$.}
\end{small}
\end{table}

\section{Model and Results}
\subsection{The contribution of Dark matter annihilation and Pulsars}

From the PAMELA and Fermi-LAT data of the electrons \cite{pamela,Fermi2012},
we can see that there may be a tiny excess above $\sim 100$ GeV. The ATIC
and Fermi-LAT spectra of the total electrons and positrons suggest that
there is a significant excess at energies above 300 GeV. Therefore we
adopt a three-component electron model to fit the data. The background
electrons from primary cosmic ray sources contribute to the electrons
below $\sim50$ GeV, the pulsar component to reproduce the high energy
excesses of the $e^+e^-$ spectra since pulsars are a kind of feasible
high energy positron and electron source that has been widely discussed
in literature (e.g., \cite{zhangli,fanyz}), and the annihilation of DM
particles with mass $\sim 130$ GeV.

The source function of electrons and positrons from DM annihilation is
\begin{equation}
q(E,r)=\frac{\langle\sigma v\rangle_{\rm \chi\chi\rightarrow e^{+}e^{-}}}{2m_{\chi}^2}\frac{{\rm
d}N}{{\rm d}E}\times \rho^2(r),
\end{equation}
where $m_{\chi}$ is the particle mass of DM, $\rho(r)$ is the spatial
distribution of energy density, and ${\rm d}N/{\rm d}E$ is the electron
and positron yield spectrum produced by one pair of DM annihilation.
In this work we use the Einasto DM density profile \cite{navarro}
\begin{equation}
\rho(r)=\rho_{*}\exp\left(-\frac{2}{\alpha_*}\left[\left(\frac{r}
{r_{*}}\right)^{\alpha_*}-1\right]\right),
\end{equation}
where $\alpha_*=0.17$, $r_{*}\approx 15.7$ kpc and $\rho_{*}\approx 0.14$
GeV cm$^{-3}$.

\begin{figure*}
\includegraphics[width=0.45\textwidth]{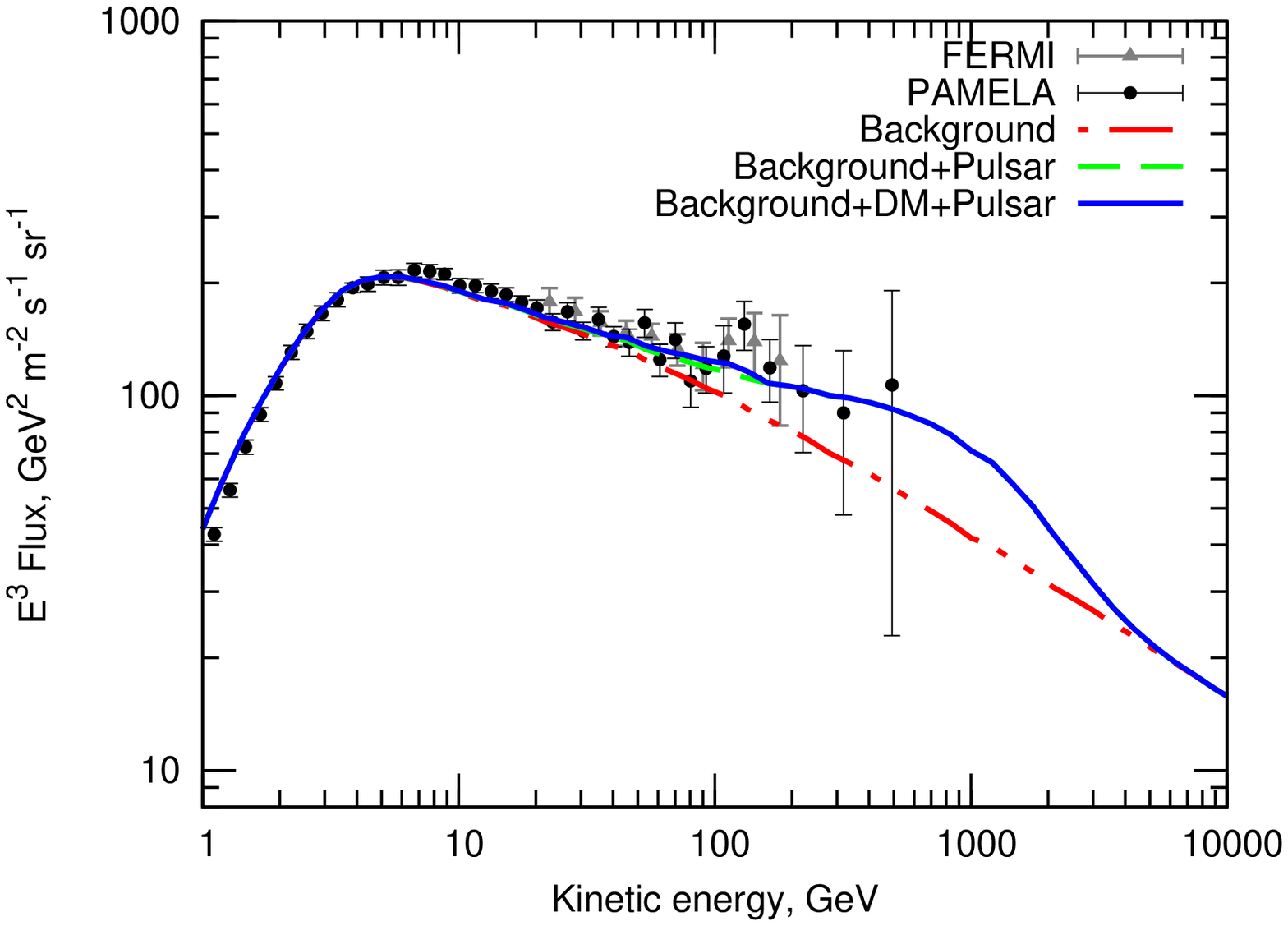}
\includegraphics[width=0.45\textwidth]{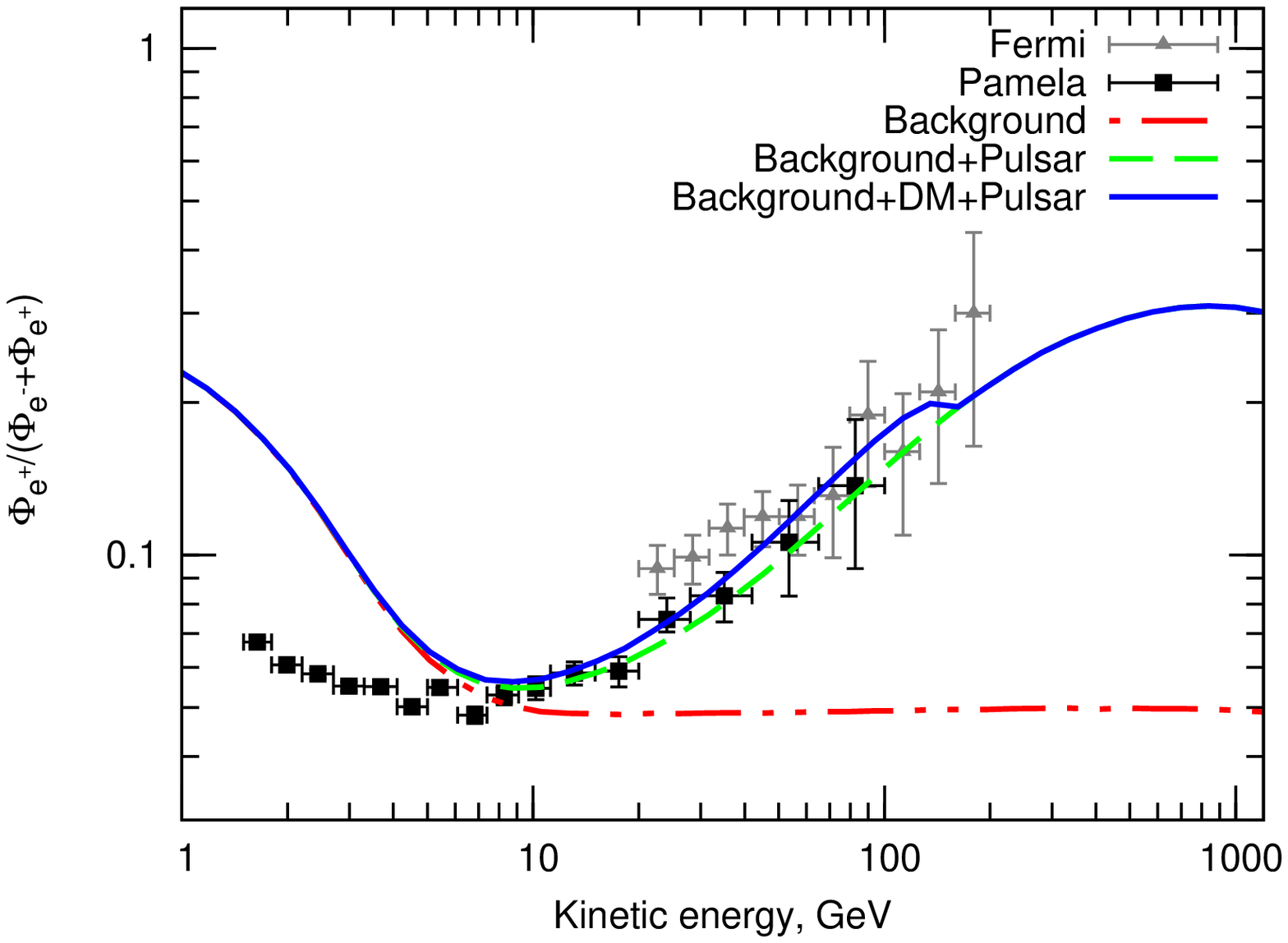}
\caption{The $e^{-}$ flux (left) and positron fraction (right) for
DR propagation model. The dash-dotted (red) line is the CR background
component, the dashed (green) line represents the sum of background
and pulsar components, and the solid (blue) line is the sum of the
above two and an additional DM component with mass $\sim130$ GeV.
The annihilation cross section of the DM is
$3.9\times10^{-26}~{\rm cm^3~s^{-1}}$. As shown in the right panel, due to lower
flux of the corresponding ``background", it will be easier to identify
a DM-origin positron component than the electron component.
References of the data are: PAMELA \cite{pamela,pamela-positron},
Fermi-LAT \cite{Fermi2012}.}
\label{fig:dr}
\end{figure*}

We assume that the high energy electrons/positrons are generated through
the cascade of electrons accelerated in the magnetosphere of pulsars
\cite{zhangli,index}. The energy spectrum of $e^+e^-$ injected to
the galaxy from pulsars can be parameterized {\bf as a broken power-law}
with the cutoff at $E_c$, ${\rm d}N/{\rm d}E \propto A_{\rm psr}
E^{-\alpha}\exp(-E/E_c)$, where $E_c$ ranges from several tens GeV
to higher than TeV, according to the models and parameters of the
pulsars \cite{zhangli,Malyshev}. And the power-law index $\alpha$
ranges from 1 to 2.2 depending on the gamma-ray and radio
observations \cite{index}. In this Letter, we adopt $E_c= 860$
GeV and $\alpha=1.28$ following \cite{liujie}. The spatial distribution
of pulsars can be parameterized the following form \cite{distrib}
\begin{equation}
f(R,z)\propto\left(\frac{R}{R_{\odot}}\right)^a\exp\left[-\frac{b(R-
R_{\odot})}{R_{\odot}}\right]\exp\left(-\frac{|z|}{z_s}\right),
\end{equation}
where $R_{\odot}=8.5$ kpc is the distance of solar system from the
Galactic center, $z_s\approx 0.2$ kpc is the scale height of the
pulsar distribution, $a=2.35$ and $b=5.56$. The other parameter
appear in the above equation is the normalization factor $A_{\rm
psr}$, and it will be determined in our modeling.

\begin{figure*}
\includegraphics[width=0.45\textwidth]{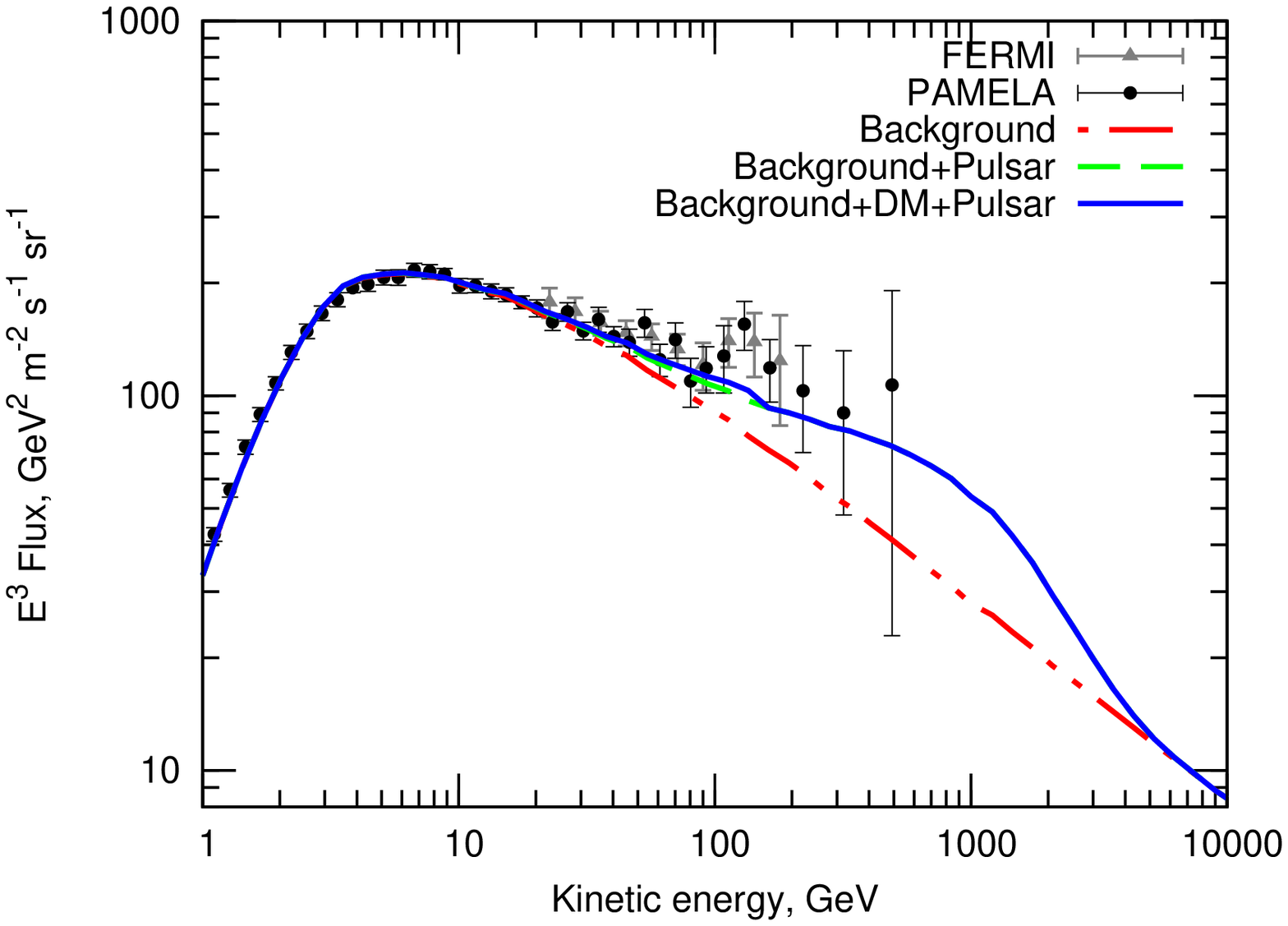}
\includegraphics[width=0.45\textwidth]{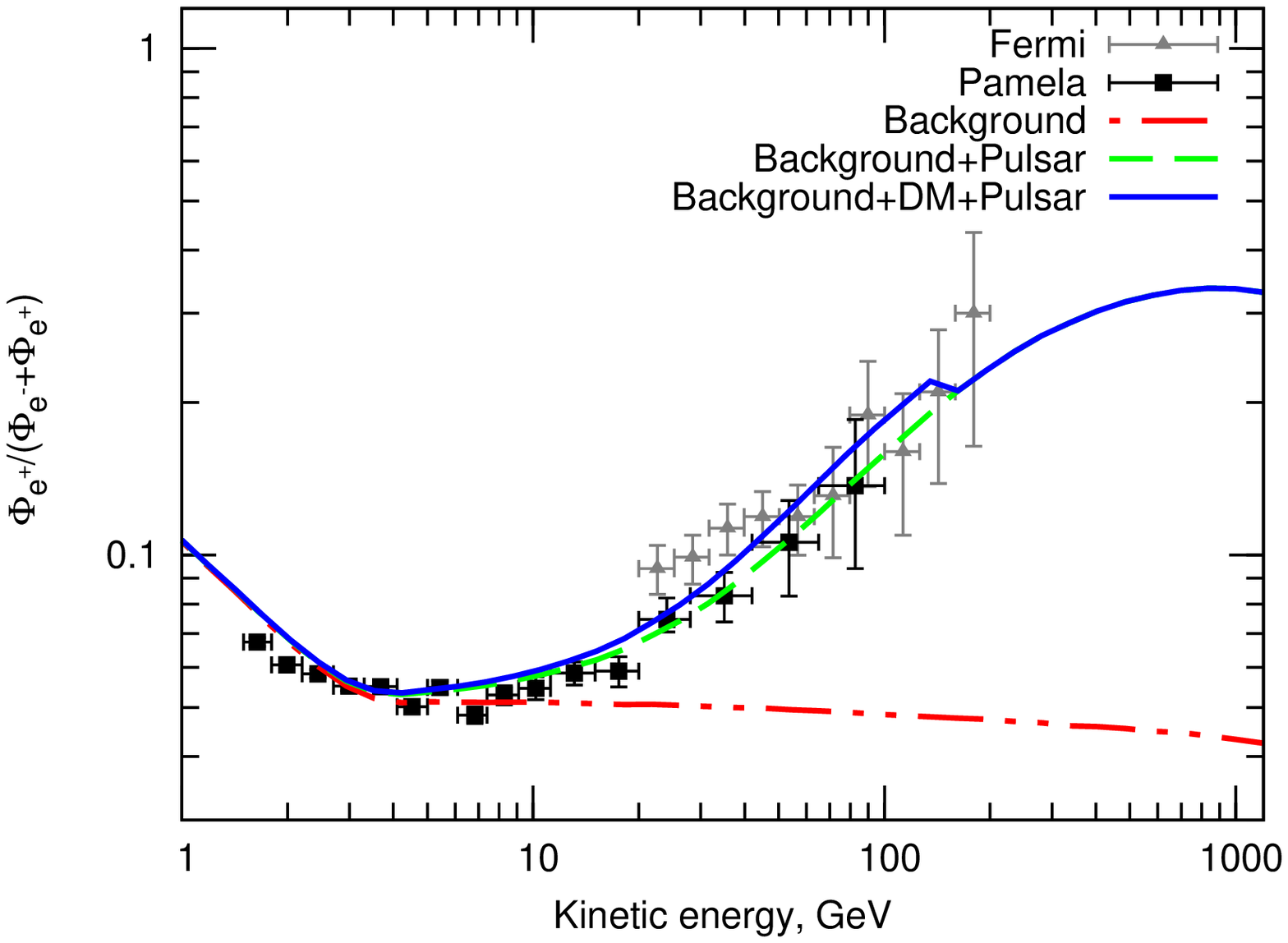}
\caption{ Same as Figure \ref{fig:dr} but for the DC propagation model.
The DM annihilation cross section is
$4.0\times10^{-26}~ {\rm cm^3~s^{-1}}$.}
\label{fig:dc}
\end{figure*}

\subsection{The Results}

Figure \ref{fig:dr} shows the results of the three-component model
in the DR propagation model. The CR electron
background is calculated using the parameters in Table \ref{table1}.
Note for the background positron (and the secondary electron) flux we
multiply a constant factor $c_{e^+}=1.4$ to better fit the data
\cite{liujie}, which may account for the uncertainties of the propagation
model, interstellar gas distribution and the inelastic hadronic interaction
model. To judge the improvement of the fit through adding the
$\sim 130~{\rm GeV}$ DM component, we calculate the $\chi^2$ value of
the model. To minimize the effect of solar modulation, we use the data
with energies higher than $10$ GeV for $\chi^2$ calculation, although we
employ a force field approximation \cite{gleeson} to approach the solar
modulation with modulation potential $\sim400$ MV. The $\chi^2$
for the background $+$ pulsar model (null hypothesis) is $20.9$.
With the presence of a $\sim 130$ GeV DM component, the minimum $\chi^2$
found is $18.9$, and the best-fit cross section is $\langle\sigma v
\rangle_{\chi\chi\rightarrow e^{+}e^{-}}={\rm 3.9\times 10^{-26}~
cm^3~s^{-1}}$. The fit is slightly improved but not significantly enough
in case of an additional degree of freedom. The small difference between
the $\chi^2$ inferred in these two scenarios suggests that the background
$+$ pulsar model is enough to describe the data.

Figure \ref{fig:dc} is the same as Figure \ref{fig:dr} except
that it is for the diffusion-convection propagation model.
Similarly we calculate the likelihood of the DM component.
The $\chi^2$ for the null hypothesis is $21.5$, while it is $19.9$ in the
presence of a $130$ GeV DM component. The best fit cross section is
$4.0\times 10^{-26}$ cm$^3$ s$^{-1}$. As in the DR model, the background
plus pulsar model gives reasonable fit to the data.

\begin{figure*}
\begin{center}
\includegraphics[width=0.6\textwidth]{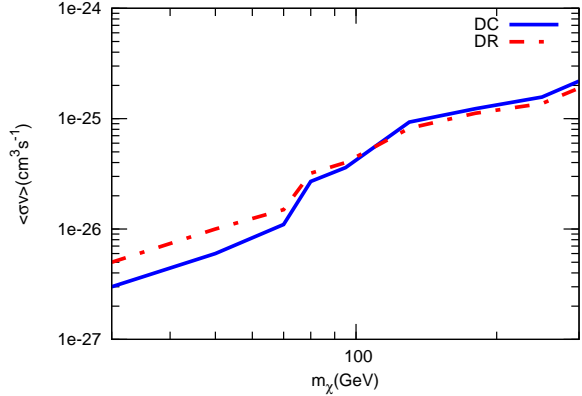}
\caption{The 2$\sigma$ upper limits on $\langle\sigma v\rangle
_{\chi\chi\rightarrow e^{+}e^{-}}$ for DR (dash-dotted)
and DC (solid) propagation models respectively.}
\label{fig:xianzhi}
\end{center}
\end{figure*}

It is shown above that adding a DM component do not significantly improve the fitting, so we turn to set an upper limit of the DM
component instead. The $95\%$ confidence level upper limits of the
annihilation cross section $\langle\sigma v\rangle_{\chi\chi\rightarrow
e^{+}e^{-}}$ for different mass $m_{\chi}$ are shown in Figure
\ref{fig:xianzhi}. For $m_\chi=130$ GeV, the 2$\sigma$ upper limit
of the cross section is about $8\times10^{-26}$ cm$^3$ s$^{-1}$.
This constraint is stronger than that derived through $\gamma$-ray
observations of the Milky Way halo \cite{fermiline}. One caution is that the constraints also depend on the assumptions used for the background, which are not same as in \cite{fermiline}.

Since the current observational data are not precise enough to see
whether there is a tiny structure of the electron spectra at $\sim 100$
GeV, we would like to discuss the potential of the AMS-02 experiment on
this issue. We use Monte Carlo simulation to produce the expected electron
spectra of AMS-02, based on the theoretical fluxes of electrons $\phi_e$
for the background $+$ pulsar $+$ DM scenario as given in Figures
\ref{fig:dr} and \ref{fig:dc}. The events number of AMS-02 can be
estimated as \cite{pato}
\begin{equation}
N=\Delta t~ A_e ~\int_{\Delta E} {\rm d}E~\int {\rm d}E^{\prime}~
\phi_e(E^{\prime})~\frac{1}{\sqrt{2\pi}\sigma}e^{-\frac{(E^{\prime}-E)^2}
{2\sigma^2}},
\end{equation}
where $\sigma=\sqrt{\left(0.106/\sqrt{E/{\rm GeV}}\right)^2+0.0125^2}
\times E/2$ \cite{energy} represent the energy resolution, $\Delta t$ is the
operating time, $\phi_e$ is the electron flux, and $A_e$ is the geometrical
acceptance of electron which is taken to be $\rm 0.045~m^2~sr$ ~\cite{app}.
The electrons are binned into $25$ bins logarithmically from $10$ GeV
to $1$ TeV. The observed number of events in each energy bin is generated
based on a Possion distribution with expected value $N$ calculated by Eq.
(5). The logarithm of the likelihood function is defined as
\begin{equation}
\ln\mathcal{L}=\sum_i N_i\ln \phi_e^i-\phi_e^i-\ln N_i!,
\end{equation}
where $N_i$ is the simulated observational electron counts in energy bin
$i$, and $\phi_e^i$ is the expected counts in the energy bin. The {\it
test statistic} of the DM signal is defined as ${\rm TS}=-2\ln(\mathcal{L}
_{\rm null}/\mathcal{L}_{\rm best})$, where $\mathcal{L}_{\rm null}$
is the best fit likelihood of null hypothesis (background $+$ pulsar
scenario), and $\mathcal{L}_{\rm best}$ is the best fit likelihood
of the model with 130 GeV DM component. For a series of annihilation
cross sections, we calculate the TS value and find the required
exposure time $T$ which makes ${\rm TS\approx25}$ (approximately
corresponding to $5\sigma$ significance). The results are summarized
in Table \ref{table2}. It is shown that if the DM annihilation cross section
is larger than $10^{-26}~{\rm cm^3~s^{-1}}$, this $\rm ~130~ GeV$ DM
signal may be identified by AMS-02.

\begin{table}[!htb]
\begin{center}
\caption {The predicted operating time for AMS-02 to identify the 130 GeV
DM annihilating into $e^+e^-$.}
\begin{tabular}{ccccccc}
\hline \hline
 & $\langle\sigma v\rangle_{\rm \chi\chi\rightarrow e^{+}e^{-}}$ ($\rm 10^{-26}~cm^3~ s^{-1}$) & 1.0 & 2.0 & 3.0 & 4.0 & 5.0 \\
  \hline
DR & $T$(yr) & 7.7 & 2.2  & 1.1 & 0.7 & 0.5\\
DC & $T$(yr) & 10.0 & 2.6  & 1.3 & 0.8 & 0.5\\
  \hline
  \hline
\end{tabular}
\label{table2}
\end{center}
\end{table}

We may further note that the DM signal, if exists, will be more
prominent in the positron fraction than in the electron spectrum.
As shown in Figures \ref{fig:dr} and \ref{fig:dc}, the relative
excess of the DM component is only $\sim2\%$ compared with the background
$+$ pulsar flux. However, for the positron fraction the contribution
of the DM component could be more than $10\%$ (relative to the background
$+$ pulsar components) for the best fit cross section $4\times10^{26}$
cm$^3$ s$^{-1}$.

\subsection{Constraints from other observations}

The observations of $\gamma$-rays (the internal bremsstrahlung and
inverse Compton radiation component) and/or radio emission (the synchrotron
radiation component) may constrain the current scenario that DM annihilates
into electrons/positrons \cite{constrain,fermiline}. The lack of evident
continual spectrum component associated with the 130 GeV $\gamma$-ray
line from the Galactic center suggests that the DM particles mainly
annihilate into final states with few $\gamma$-rays (such as $e^{+}e^{-}$,
$\mu^{+}\mu^{-}$, or neutrinos), otherwise the observed relic density can
not be explained \cite{Buckley2012}. Such a speculation is one of the main
motivation of this work. It was shown that the latest $\gamma$-ray
observations by Fermi-LAT can constrain the $\langle\sigma v\rangle_{\rm
\chi\chi\rightarrow e^{+}e^{-}}$ to the level of $10^{-24}$ cm$^{3}$
s$^{-1}$ for DM mass $\sim 100$ GeV \cite{fermiline}, which is much
larger than the constraint (i.e., $\langle\sigma v\rangle_{\rm
\chi\chi\rightarrow e^{+}e^{-}}<10^{-25}~{\rm cm^{3}~s^{-1}}$) yielded
in this work.

\section{Conclusion}

The spectra of the CR electrons detected by PAMELA and Fermi-LAT both
show small wiggle-like structure at $\sim 100$ GeV, potentially
consisting of a weak signal of $\sim 130$ GeV DM particles annihilating
into electrons/positrons with a cross section
$\langle\sigma v\rangle_{\rm \chi\chi\rightarrow e^{+}e^{-}} <
10^{-25}~{\rm cm^{3}~s^{-1}}$. It maybe connect with
the recently reported $\sim130$ GeV $\gamma$-ray line emission which may be the
result of DM annihilation in the Milky Way. We investigate the
contribution to the electron spectrum and positron fraction of such
a DM component. It is found that adding the $130$ GeV DM component
can improve the fit to the data, but the improvement is not
significant enough. We further use the current data to set constraints
on the DM annihilation cross section to $e^+e^-$.

As found in our modeling, the DM-origin electrons/positrons, if there
are, are likely less than $\sim 2\%$ ($4\%$) of the background $+$ pulsar
$e^{-}$ ($e^{-}+e^{+}$) flux at $\sim 130$ GeV. Hence, accurate
measurements of the spectrum of the electrons (and positrons) by AMS-02
and DAMPE/CALET\footnote{DAMPE and CALET
(http://calet.phys.lsu.edu/) are able to measure the total spectrum
of cosmic ray electrons and positrons accurately.} are highly necessary
to test the DM origin of the wiggle-like structure (see Table 2 for the
expected performance of AMS-02 in order to identify such a component).
Finally we would like to point out that it may be easier to identify
a DM-origin positron component than the electron component due to lower
flux of the corresponding ``background".

\section*{Acknowledgments}
We are grateful to the anonymous referee for the insightful comments
that help us to improve the manuscript significantly. This work is
supported in part by the 973 Program of China (No. 2013CB837000), 100
Talents program of Chinese Academy of Sciences, National Natural Science
Foundation of China (No. 11105155), Foundation for Distinguished Young
Scholars of Jiangsu Province, China (No. BK2012047), and the China
Postdoctoral Science Foundation (No. 2012M521136). QY acknowledges
the support from the Key Laboratory of Dark Matter and Space Astronomy
of Chinese Academy of Sciences.\\

\end{document}